\documentstyle[psfig,aps]{revtex}

\begin{document}

\title{A quantum field theoretic description of the delayed choice experiment}
\author{R. Srikanth  \\ 
Indian Institute of Astrophysics, \\ 
Koramangala, Bangalore- 34, Karnataka, India}
\maketitle
\date{}

\begin{abstract}
Wheeler's delayed choice experiment, a well known manifestation of 
the complementarity principle, has proved somewhat difficult to physically
interpret. We show that, restated in quantum field theoretic language, 
the experiment submits to a simple explanation: that wave- or particle-nature
is imposed not at the slit plane but at the detector system.
The intepretational difficulty conventionally encountered is due to the
assumption of enforcement of complementarity at the former.

\end{abstract}

 ~\\
{\em Introduction}:
As is well known, quantum mechanics manifests several non-classical 
phenomena arising because of superposition and entanglement \cite{gren}. 
One such is the delayed choice experiment (DCE) \cite{wheeler},
which is essentially
a dramatic realization of Bohr's complementarity principle (CP) \cite{bohr}, 
an interpretation of wave-particle duality of matter. In Bohr's viewpoint,
objective reality is denied and what we observe depends on how we ask. Only
through the irreversible act of amplification induced by measurement do
phenomena come to exist. Wheeler picturesquely allegorized Bohr's viewpoint as 
a ``smoky dragon" with its tail in the light source and mouth biting the
detector. As an illustration of CP, let's consider a
double-slit illuminated by a coherent source. An
observer behind the slit plane is equipped with a dual detector
system whereby he can
observe the diffracted light with a screen or with two
telescopes, one focussed on each slit. Detection at the screen produces a
Young's double-slit pattern as each photon passes through both slits
and interferes with itself. On the other hand, a detection at a telescope
would imply the passage of the photon through that slit on which the telescope
is focussed. This forces particle nature on the light and 
no Young's interference pattern is seen. 

In DCE, the observer
waits until after the light has passed the slit plane to decide whether he
measures the wave- or particle-nature of the light. In the popular and
scientific literature, it has provoked intriguing questions \cite{horgan} like:
how does the light ``decide" whether to pass through both slits
or one of them in order to conform to the future decision of the observer?
Does it do so via a backward-time effect? Or does it 
``know" what the observer will decide later on? 
In the following account, we present, using the formalism of quantum field 
theory, a conventional explanation of DCE in which 
this difficulty in physically interpreting the effect does not appear.
This conclusion is in agreement with Wheeler's observation that 
the aspect of matter manifested depends on the registering
device chosen \cite{whe94,home}.
Consideration of such issues in the foundations of quantum mechanics
is relevant to the burgeoning field of quantum information \cite{preskill},
because they help to 
better visualize the nature and flow of information in quantum systems.

~\\
{\em Derivation}:
We specialize to the single particle case the more general formalism used in 
the treatment of multiparticle interference \cite{strekalov}.
Consider a source S that illuminates a diaphragm O perforated by two slits
$a$ and $b$. The state of the photon is given by the two-mode state
\begin{equation}
\label{state}
|\Psi\rangle = |{\rm vac}\rangle + 
\frac{\epsilon}{\sqrt{2}}(\hat{a}^{\dag}e^{i\phi_A} + 
		\hat{b}^{\dag}e^{i\phi_B})|{\rm vac}\rangle ,
\end{equation}
where $\epsilon$ determines the strength of the optical field,
$\hat{X}^{\dag}$ is the creation operator 
for the photon mode corresponding to slit $X$, $\phi_X$ is the phase factor 
associated with the mode $X$, $|{\rm vac}\rangle$ is the underlying 
vacuum state. Hence, $|\Psi\rangle$ is a superposition of Fock states in modes
$a$ and $b$.
 
The light diffracts at the double-slit and falls on the screen to form an 
interference pattern, or perhaps on the aperture of a telescope to permit 
a path detection. The positive frequency part of the field operator at some 
point $x$ on the detector is given by 
\begin{equation}
E^{(+)}(x) = \hat{a}\exp(ik[d_{s} + d_{ax}]) +
\hat{b} \exp(ik[d_{s} + d_{bx}]) ,
\end{equation}
where $\hat{X}$ is the annihilation operator for the mode corresponding
to slit $X$, $k$ is the wave-number, $d_{s}$ the distance from the coherent
source to either slit and $d_{Xx}$
the distance from slit $X$ to point $x$ on the screen. The probability 
$P(x)$ for detecting a photon at point $x$ on the screen is given by 
$\langle E^{(-)}(x)E^{(+)}(x) \rangle$,
where the angles 
$\langle \cdot\cdot\cdot \rangle$ represent expectation value in the state
$|\Psi\rangle$. We find
\begin{equation}
\label{probscre}
P(x) \propto 1 + \cos(d_{ax} - d_{bx}) ,
\end{equation}
if we set $\phi_A = \phi_B$, though this is not necessary to observe 
fringes. Eq. (\ref{probscre}) is the usual far-field 
Young's double-slit pattern. 

On the other hand, let us suppose
 the screen is replaced with a telescope focussed on one of the slits, say
$X$. A detection with it is represented by the positive frequency part of 
the field operator, $E^{(+)}_X$,
whose measurement implies the photon's passage through slit $X$. Then
$E^{(+)}(x)$ is given by
\begin{equation}
E^{(+)}_a(x) \equiv \hat{a}\exp (ik[d_{s} + d_{a\xi}]) \hspace{1.0cm}
{\rm or}   \hspace{1.0cm}
E^{(+)}_b(x) \equiv \hat{b}\exp (ik[d_{s} + d_{b\eta}]). 
\end{equation}
Here $x = \xi$ is the
position of the telescope focussed on slit $a$, and $x = \eta$, that of
the telescope focussed on slit
$b$. The probability for detection at either telescope is given by
$\langle E^{(-)}_a(x)E^{(+)}_a(x) \rangle = \langle E^{(-)}_b(x)E^{(+)}_b(x)
\rangle \propto \epsilon^2$.
Therefore in this case, the probability for detection on either telescope is
uniform and shows no fringes. This, as well as the result Eq.
(\ref{probscre}), is in keeping with what one expects for unentangled 
quantum systems on basis of the
complementarity principle: that path information and first-order interference 
pattern are mutually exclusive.

We note that there is no explicit reference to time in the above calculation. 
Thus, there is no reason to expect that these results
should not hold if the observer chooses to use the screen or the telescopes
after the light crosses the slit plane. 
Therefore, the results derived above are sufficient to
explain how delayed choice works. 
 
~\\
{\em Physical interpretation}:
Let us physically interpret the above results. The main point is that CP is
not enforced on the photons 
at the slit plane. The photon passing through the slits
does not need to ``decide" whether to pass through both slits or one of them.
It passes through both, irrespective of whether the observer subsequently 
measures position or momentum. The decision to manifest particle
or wave nature occurs at the detector system according the observable, and
hence allowed eigenstates, chosen by the observer.
Amplitude information from both paths superpose at all $x$'s. Even when the
observer trains his telescope on one of the slits, amplitude information
from both slits fall on the telescope aperture. But subsequently, the 
amplitude for one of the slits is filtered out by the telescope
optics until only that
for the focussed slit falls on the eye-piece. This, of course, is equivalent
to measurement with one of the operators $E^{(+)}_X$. 

Similar arguments apply
also to a quantum eraser in which paths are potentially distinguishable
via entanglement
or one of the paths is unitarily marked (say, by a polarization rotator)
\cite{scully}.
This would force the expression of particle nature.
However, as in DCE, the total information is not destroyed but remains
hidden.  Subsequent erasure of path information
re-manifests correlated interference based on the superposition of
amplitudes from both paths. 
However, we note that the status of complementarity is not always obvious in
entangled systems. For example, a possible non-standard effect is discussed
in Refs. \cite{srik,altschul}. 
 
The assumption implicit in DCE that in order to 
manifest particle nature the photon should have passed through only
one of the slits turns out to be unnecessary for obtaining the required
statistical predictions of the theory. It is this feature that
frees the current explanation from having to invoke backward-time effects
or cognitive interpretations. Suppose on the other hand that the slits are
equipped with a suitable device to find out the slit the photon passes
through. This, of course, forces a particle-like behaviour and destroys
interference in the traditional sense of the complementarity principle.
In this case there is a genuine lack of simultaneous
information from both slits. Thus, we see that complementarity can be enforced
both by an intermediary which-path measurement as well as at the final 
detection. 
These considerations suggest that
wave-particle duality reflects a deeper information theoretic 
{\em and} quantum field theoretic nature of photons and, in general, matter.


\end{document}